\documentstyle[prl,aps,epsfig,array,hhline,twocolumn,amsmath,amssymb]{revtex}
\input psfig.sty

\begin{document}
\draft

\title{A Discrete Four Stroke Quantum Heat Engine 
Exploring the Origin of Friction.}

\author{Ronnie Kosloff and Tova Feldmann}

\address{
Department of Physical Chemistry 
the Hebrew University, Jerusalem 91904, Israel\\
}

\maketitle
\begin{abstract}
The optimal power performance of a first principle quantum heat engine model shows
friction-like phenomena when 
the internal fluid Hamiltonian does not commute with   the external 
control field. The model is based on interacting two-level-systems
where the external magnetic field serves as a control variable.

\end{abstract}

\section{Introduction}
\label{sec:introduction}

It is well established that the performance of working heat engines are limited 
by intrinsic unavoidable irreversibilities. Maximum power is obtained at 
the expense of efficiency where the reversible point of maximum efficiency has zero power.
This principle has been  clearly illustrated by
the endoreversible model of Curzon and Ahlborn \cite{curzon75} and summarized 
by Salamon et. al. \cite{salamon2001}. Two additional unavoidable 
sources of loss are  heat leaks which practically eliminate
the maximum efficiency adiabatic operation, and internal friction 
which restricts fast operating cycles.

Is this universal performance limitation of heat engines macroscopic or microscopic?
Though the common image of heat engines is of large  macroscopic devices, 
microscopic models based on first principle quantum mechanics
are limited by the Carnot efficiency \cite{geusic59}, 
and show a remarkable resemblance to their macroscopic analogs 
when the engines produce finite power \cite{k24}. 

In previous studies, both discrete \cite{geva0,geva1,feldmann96,lloyd,feldmann00}
and continuous quantum models \cite{k24,geva2,k156,jose01} have been scrutinized,
which are analogous, respectively, to four-stroke-engines and turbines. 
The present study examines a discrete four-stroke quantum  engine, comparing it
to an engine subject to phenomenological internal friction.
It will be demonstrated that the
quantum engines inability to control simultaneously the external and internal portions
of the working fluid Hamiltonian is its source of friction.

\section{Basic construction}
\label{sec:engine}

All working heat engines and all refrigerators operate on the same principle.
The engine manipulates the energy flow between three reservoirs, a hot,
cold and power
reservoir, either to extract power from the temperature difference
or to pump heat from the cold to the hot reservoir at the expense
of external power. 

The present model of a discrete quantum  heat engine is composed
of a cycle of operation constructed from two {\em adiabats} and two 
{\em isotherms} similar to a Otto cycle.
The quantum dynamics  are generated by
external fields in the {\em adiabats} and by heat flows from hot and cold
reservoirs in the {\em isochores}.
The working medium is modeled as a gas of interacting particles with the
Hamiltonian: 
\begin{equation}
{\bf \hat H} ~~=~~  {\bf \hat H_{ext}} ~+~  {\bf \hat H_{int}}~~.
\label{eq:hamil}
\end{equation} 
${\bf \hat H_{ext}}= \omega \sum_i {\bf \hat H_i} $ is the sum 
of single particle Hamiltonians, $\omega= \omega(t)$ 
is the time dependent external control field, and $ {\bf \hat H_{int}}$
represents the inter-particle interaction.

The change in time of an operator ${\bf \hat A}$
during the {\em adiabatic/isochoreal} branches is described as:
\begin{equation}
\dot {\bf \hat A}~~=~~ i [{\bf \hat H}, {\bf \hat A}] ~+~ \frac{\partial {\bf \hat A}}
{\partial t}~~+~~{\cal L}_{D}({\bf \hat A})~~~,
\label{eq:isodyn}
\end{equation} 
${\cal L}_{D} $ represents the Liouville dissipative generator on the {\em isochore}
in contact  with either the hot or cold bath (units of $\hbar=1$).
Replacing ${\bf \hat A}$ by ${\bf \hat H}$ in Eq. (\ref{eq:isodyn})
{leads to the power invested in or extracted from the {\em adiabats}:,
\begin{equation}
{\cal P} ~~=~~ \dot \omega \sum_i \langle  {\bf \hat H_i} \rangle~~~,
\label{eq:power}
\end{equation}
where $\langle  {\bf \hat H_i} \rangle$ is the expectation value of the
single particle Hamiltonian.
The heat flow is extracted from the energy balance on the 
{\em isochores} \cite{k24,feldmann00}:
\begin{equation}
\dot {\cal Q}_{h/c}~~=~~ \langle {\cal L}_{h/c} \left( {\bf \hat H} \right)\rangle~~.
\label{eq:heatflow}
\end{equation}

For simplicity,
the single particle Hamiltonian is chosen as a two-level-system (TLS):
${\bf \hat H}_i={\boldsymbol{\mathrm{\hat{\sigma}}}}_z^i$. 
The interaction term is restricted to coupling of pairs of spin atoms.
As a result, the state of the working medium is described by
an ensemble of pairs of two-level-systems
represented by the density operator $\hat \rho$,
defined in the tensor product space of the individual two TLS systems. 
Expectation values are obtained by the usual definition 
$ \langle  {\bf \hat A} \rangle = tr \{  {\bf \hat A}\hat \rho \} $.
The external Hamiltonian then becomes:
\begin{equation}
{\bf \hat H_{ext}} ~~=~~ \omega \left({\boldsymbol{\mathrm{\hat{\sigma}}}}_z^1 
\otimes {\bf \hat I^2}
+
{\bf \hat I^1} \otimes \hat \sigma_z^2  \right)
\end{equation}
and the external field is chosen to be in the  $z$ direction. 
The interaction Hamiltonian is chosen as:
\begin{equation}
{\bf \hat H_{int}} ~~=~~ J \left({ {\boldsymbol{\mathrm{\hat{\sigma}}}}_x^1} \otimes { {\boldsymbol{\mathrm{\hat{\sigma}}}}_x^2} -
{{\boldsymbol{\mathrm{\hat{\sigma}}}}_y^1}\otimes {\boldsymbol{\mathrm{\hat{\sigma}}}}_y^2 
  \right)~~.
\label{eq:interaction}
\end{equation}
$J$ scales the strength of the interaction. When $J \rightarrow 0$ the 
model approaches the previously studied frictionless model \cite{feldmann96}.
The inter-particle interaction term, Eq. (\ref{eq:interaction}), defines a correlation
energy between the two single particle spins in the $\vec x$ and $\vec y$ direction. 
As a result, $[{\bf \hat H_{ext}},{\bf \hat H_{int}}] \ne 0$, 
since the external Hamiltonian is polarized in the $\vec z$ direction.

\section{Dynamics of the working medium}
\label{subsec:dynamics}

The dynamics generated by the Hamiltonian is completely
determined by the algebra of commutation relations of
the set of 16 operators spanning the Hilbert space of the combined system.
Due to symmetry, the commutation algebra decomposes into subsets of 
operators with closed commutation relations. The set is generated by
commutation relations between the operators composing the Hamiltonian:
\begin{eqnarray}
\nonumber
\begin{array}{c}
{\bf \hat B}_{1}
\end{array} =&~~~
\begin{array}{c}
{\boldsymbol{\mathrm{\hat{\sigma}}}}_z^1 \otimes {\bf \hat I^2}
~+~
{\bf \hat I^1} \otimes {\boldsymbol{\mathrm{\hat{\sigma}}}}_z^2  
\end{array} 
~~~~,~~~~\\
\begin{array}{c}
{\bf \hat B}_{2}
\end{array} =&
\begin{array}{c}
{{\boldsymbol{\mathrm{\hat{\sigma}}}}_x^1} \otimes {{\boldsymbol{\mathrm{\hat{\sigma}}}}_x^2} -
{{\boldsymbol{\mathrm{\hat{\sigma}}}}_y^1} \otimes {{\boldsymbol{\mathrm{\hat{\sigma}}}}_y^2}
\end{array} 
\label{matB2}~~~~.
\end{eqnarray}
The commutation relation:
$ [{\bf \hat B}_{1},{\bf \hat B}_{2}] = 4 i {\bf \hat B}_{3}$
leads to the definition of ${\bf \hat B}_{3}$ which closes the set i.e.
the set of operators ${\bf \hat B}_{1},{\bf \hat B}_{2}, {\bf \hat B}_{3}$
forms a closed subset of the Lie algebra of the combined system.
The Hamiltonian expressed in terms of the set of operators becomes:
${ \bf \hat H }~=~ \omega {\bf \hat B_{1}}+\rm J {\bf \hat B_{2}} $

The commutation relations of the set of  $\bf \hat B_i$ operators 
are isomorphic to the angular momentum commutation relations when
the transformation $\frac{1}{4}{\bf \hat B_{\rm i}}~ \rightarrow ~{\bf \hat J_{\rm i}}$
is applied. This similarity can be exploited to
express the  expectation values in a Cartesian three 
dimensional space, where the external field is in the $\bf B_1$ or 
$\vec z$ direction,  the correlations in the  $\bf B_2$ 
or $\vec x$ direction and  $\bf B_3$ is in the $\vec y$ direction.

The closed set of operators $\bf \hat B_i$ is sufficient to follow the changes in
energy and to obtain the power consumption.
Using Eq. (\ref{eq:isodyn}) and the commutation relations of the set of 
${\bf \hat B}_{\rm i}$ operators,
the Heisenberg equation of motion for this set becomes:
\begin{eqnarray}
\frac{d}{dt}\left( \begin{array}{c}
{\bf B_{1}} \\
{\bf B_{2}} \\
{\bf B_{3}} \\
\end{array} \right) ~=~ \left(
\begin{array}{ccc}
0&0&4J \\
0&0&-4 \omega(t) \\
-4J&4 \omega(t)&0 \\
\end{array} 
\right)
\left(\begin{array}{c}
~\bf B_{1} \\
~\bf B_{2} \\
~\bf B_{3} \\
\end{array} \right)
\label{eqmotVN}
\end{eqnarray}
These equations can be written in matrix form for the expectations $b_i=tr\{{\bf \hat B_i},\hat \rho\}$:
\begin{equation}
\frac{d}{dt} {\vec {\bf  b}} ~~=~~ {\cal A}(t) {\vec  {\bf b}}
\label{eq:vec}
\end{equation}
Since the matrix ${\cal A}(t)$ is time dependent,
the propagation is broken into $N$ short time segments $\Delta t$ 
where $N \Delta t = t$, and is solved numerically. 
The matrix ${\cal A}$ is diagonalized
for each time step, assuming that $\omega(t)$ is constant within the time
period $\Delta t$. The corresponding eigenvalues become:$-4i \Omega,~0~$ and $4i \Omega$,
where $\Omega = \sqrt{J^2+ \omega^2}$.
The short time propagator for the {\em adiabats} from time $t$ to $t+\Delta t$:
\begin{eqnarray}
\begin{array}{c}
{\cal U}_a(t,\Delta t)= e^{{\cal A}(t) \Delta t}
\end{array}
=\left(
\begin{array}{ccc}
{{\omega}^2+c{J^2}} \over {\Omega}^2&{{\omega}J(1-c)} 
\over {\Omega}^2
 &Js \over \Omega \\
{{\omega}J(1-c)} \over {\Omega}^2&{J^2+c{\omega}^2} 
\over {\Omega}^2 &
{{{-\omega}s} \over  \Omega}\\
 -Js \over \Omega &{{\omega}s} \over \Omega&c \\
\end{array} \right)
\label{propag}
\end{eqnarray}
where $ c=\cos({4 \Omega}\Delta t)$ and  $ s=\sin({4 \Omega}\Delta t)$.

On the isochores, the system is in contact with a thermal bath
which eventually will lead the working fluid to thermal equilibrium with temperature $T$:
\begin{equation}
{\hat \rho}_{eq}~~=~~\frac{e^{-\beta{\bf \hat H}}}{Z}
\label{eq:thermal}
\end{equation}
with $\beta = 1/k_b T$ and $Z~=~tr\{e^{-\beta{\bf \hat H}}\}$. 
The dynamics generated by the  system-bath interaction is described by the 
dissipative Liouville operator ${\cal L}_D$, which in Lindblad 
form becomes \cite{lindblad76}:
\begin{eqnarray}
{\cal L}_{D} (\bf \hat X)  ~~=~~
\Large 
\sum_{i}~
{\bf \hat F}_{\rm i} 
{\bf \hat X} {\bf \hat F}_{\rm i}^{\dagger} -  { 1 \over 2 } ( \bf \hat F_{\rm i}
{\bf \hat F}_{\rm i}^{\dagger} {\bf \hat X}~+~{\bf \hat X} {\bf \hat F}_{\rm i}
{\bf \hat F}_{\rm i}^{\dagger})
\label{eq:Lindblad2}
\end{eqnarray}
\normalsize
where $\bf \hat F_{\rm i}$ are operators from the Hilbert space of the system.
The operators $\bf \hat F_{\rm i}$ which 
control the approach to thermal equilibrium become
the transition operators between the energy eigenstates.

Substituting the $\bf \hat B_i$ operators into Eq. (\ref{eq:Lindblad2}) one gets: 
\begin{eqnarray}
\begin{array}{l}
{ \cal L}_{D} ({\bf \hat B_1})  ~~=~~-\Gamma ~{\bf \hat B_1} + {2 \omega \over
\Omega} ({k \downarrow}-{k \uparrow} )~{\bf \hat I} \\
{ \cal L}_{D} ({\bf \hat B_2})  ~~=~~-\Gamma ~{\bf \hat B_2} + {2 \rm J \over
\Omega} ({k \downarrow}-{k \uparrow} )~{\bf \hat I} \\
{ \cal L}_{D} ({\bf \hat B_3})  ~~=~~-\Gamma ~{\bf \hat B_3} 
\end{array} 
\label{LindbB3}
\end{eqnarray}
where $\Gamma~=~{k \downarrow}+{k \uparrow}$, and the coefficients
${k \downarrow}$ and ${k \uparrow}$ obey detailed balance
$\frac{k \uparrow}{k \downarrow}~~=~~e^{-\beta{2 \Omega}}$, with the bath temperature 
$\beta=\frac{k_b}{ T_{h/c}}$.
The set of $\bf \hat B_i$ operators and the identity operator $\bf \hat I$
form a closed set to the application of the dissipative operator ${ \cal L}_{D}$. 

The relaxation to equilibrium is accompanied by loss of phase. Additional dephasing
can be caused by elastic bath fluctuations which modulate the systems frequencies.
This pure dephasing conserves the systems energy ${{\cal L}_D}^d ({\bf \hat H})=0$.
It is obtained by inserting the Hamiltonian in Lindblad's form ( Eq. (\ref{eq:Lindblad2}) )
as one of the operators $\bf \hat F_{\rm i}$:
\begin{equation}
{{\cal L}_D}^d ({\bf \hat X})~~=~~\gamma [\bf \hat H,[\bf \hat H,\bf \hat X]]~~,
\label{eq:dephasing}
\end{equation}
Equations of motion for the  set of $\bf \hat B_i$ operators on the {\it isochore}
are obtained from Eqs. (\ref{LindbB3}) (\ref{eq:dephasing}) and (\ref{eq:isodyn}):
\begin{equation}
\frac{d}{dt}{  \vec {\bf b}}  ~~=~~ {\cal B} {\vec {\bf  b} } + {\vec  {\bf c}}
\label{eq:vecb}
\end{equation}
where:
\begin{eqnarray}
\nonumber
\begin{array}{c}
{\cal B}
\end{array} ~=&~
\left(
\begin{array}{ccc}
-\Gamma+16 \gamma J^2 & -16 \gamma J \omega &4J \\
-16 \gamma \omega J &-\Gamma+16 \gamma \omega^2&-4 \omega \\
-4J&4 \omega&-\Gamma+16 \gamma \Omega^2 \\
\end{array} 
\right)&
~~,\\
{\vec  {\bf c}}~=&~
\left(
\begin{array}{c}
{2 \omega \over \Omega}( {{k \downarrow}-{k \uparrow}})     \\
{2 J \over \Omega}( {{k \downarrow}-{k \uparrow}})          \\
{0} \\
\end{array} \right)&  ~~.
\label{eq:cvec}
\end{eqnarray}
The solution of Eq. (\ref{eq:vecb}) for the {\it isochores} becomes:
\begin{eqnarray}
\vec {\bf b } (t)~~=~~ {\cal U}_{T}(\vec {\bf b} (0) - \vec {\bf b^{eq}}) + \vec {\bf b^{eq}}
\label{soliso}
\end{eqnarray}
where $ \vec {\bf b^{eq}} = -\frac{1}{\Gamma} \vec {\bf c}$ and
\begin{eqnarray}
\begin{array}{c}
{\cal U}_{T}
\end{array}
=e^{-(\Gamma-16 \gamma \Omega^2) t}
\left(
\begin{array}{ccc}
{X{\omega}^2+c{J^2}} \over {\Omega}^2&
{{\omega}J(X-c)} \over {\Omega}^2
 &Js \over \Omega \\
{{\omega}J(X-c)} \over {\Omega}^2&{X J^2+
c{\omega}^2} \over {\Omega}^2 &
{{{-\omega}s} \over  \Omega}\\
 -Js \over \Omega &{{\omega}s} \over \Omega& c \\
\end{array} \right)~,
\label{PropagTNew}
\end{eqnarray}
where  $X=\exp(-{16 \gamma \Omega^2 t})$, $ c=\cos({4 \Omega}t)$ and  
$ s=\sin({4 \Omega}t)$.

\section{The Cycle of Operation}
\label{subsec:cycle}

Fig. \ref{fig:cycle1} illustrates the cycle of operation 
on the plane of the variables $\omega$, the external control, 
and the projection of the polarization on the energy axis $E/\Omega$. 
A different  view is
displayed in Fig. \ref{fig:traj} showing the cycle trajectory
on the volume defined by the set of "polarization" coordinates 
$\langle \bf \hat B_i \rangle$.
\begin{figure}[tb]
\vspace{-0.66cm}
\hspace{3.cm}
\psfig{figure=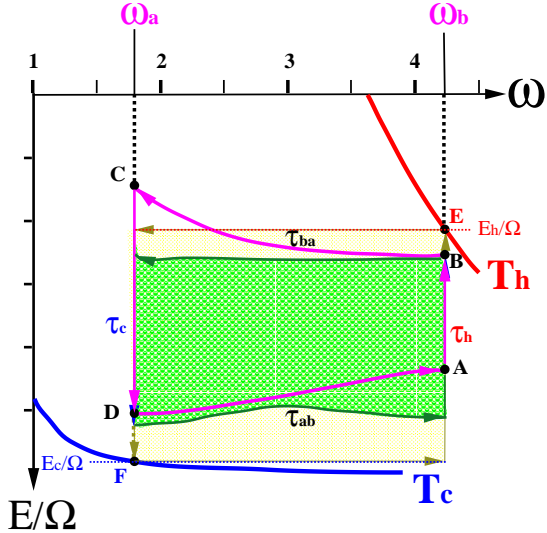,width=0.4\textwidth}
\vspace{0.1cm}
\caption{The heat engine with $J \ne 0$ in the $\omega~$, $E/\Omega$ plane.
$T_h$ is the hot  bath temperature.
$\tau_h$ is the time allocation when in contact with the hot bath.
$T_c$ and $\tau_c$ represent the temperature and time allocation
for the cold bath. $\tau_{ba}$ represents the time allocation for compression
(field change from $\omega_b$ to $\omega_a$) and $\tau_{ab}$ for expansion. 
The rectangular box which includes point $E$ on the hot {\em isochore} 
where the system is in equilibrium with $T_h$,
and point $F$ (equilibrium with $T_c$ ) on the cold {\em isochore},
represents the cycle of maximum
work. This cycle spends an infinite amount of time on  all branches.
The work is the area of the rectangle 
${\cal W}_{max}=(\Omega_b-\Omega_a)\cdot(\frac{E_h}{\Omega_b}-\frac{E_c}{\Omega_a})$.
The optimal power cycle is emphasized by the green shading.  
The time allocations for this cycle are $\tau_h=1.705$,$\tau_{ba}=0.4369$
,$\tau_c=1.76597321$ and $\tau_{ab}=0.4953$. The cycle depicted in purple is characterized
by very short time allocations on the adiabats ($\tau_{ba}=\tau_{ab}=0.00025$). 
The common engine parameters are: $\omega_a=1.794$, $\omega_b=4.239$ $T_h=2.5$, $T_c=0.5$,
$J=0.6$. $\Gamma_c=\Gamma_h=1$ with units where $\hbar=1$ and $k_b=1$.}
\label{fig:cycle1}
\end{figure}
The cycle starts at point $\bf A$ where the system is in contact 
with the hot bath at temperature $T_h$. The system accumulates heat
for a period $\tau_h$ until it reaches point $\bf B$. Point $\bf E$
is the equilibrium point of the scaled energy $E / \Omega$ at the bath temperature
$T_h$ with external field strength $\omega_b$. {\em Adiabatic} compression 
from $\omega_b$ to $\omega_a$ follows the trajectory
from point $\bf B$ to point $\bf C$ for a time duration $\tau_{ba}$
with a constant $\dot \omega$. 
The system is in contact with the cold bath from point $\bf C$
to point $\bf D$ for a time duration $\tau_c$. The cold bath equilibrium point, at
temperature $T_c$ at $\omega_a$, is $\bf F$. The cycle becomes closed by a compression
stage from point $\bf D$ back to point $\bf A$.
 \begin{figure}[tb]
\vspace{-0.66cm}
\hspace{3.cm}
\psfig{figure=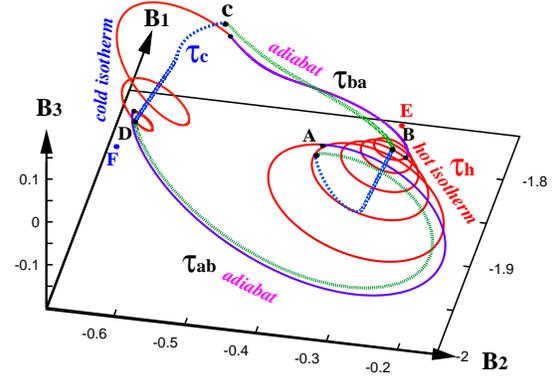,width=0.4\textwidth}
\vspace{-0.2cm}
\caption{The optimal power heat engine cycle corresponding to Fig. 1. 
The $\bf \hat B_1$ coordinate represents the individual particle 
polarization, and the $\bf \hat B_2$ and $\bf \hat B_3$ coordinates
represent the interparticle correlation.
The inner cycle with blue {\it isochores} is subject to strong dephasing
($\gamma_h=0.15 $ and $\gamma_c=0.5$).}
\label{fig:traj}
\end{figure}
For long time duration on the {\em adiabats} the cycle of operation
is restricted to the $\bf \hat B_1$, $\bf \hat B_2$ plane. For fast motion
on the {\em adiabats} the system cannot follow the instantaneous change
in the direction of the Hamiltonian  rotating on the 
$\bf \hat B_1$, $\bf \hat B_2$ plane. As a result the expectation of
the $\bf \hat B_3$ operator increases starting a precession type motion 
around the temporary direction of the Hamiltonian 
${\bf \hat  H}=\omega(t){ \bf \hat B_1}+J {\bf \hat B_2}$. 
This can be seen clearly in Fig.  \ref{fig:traj} in the  trajectory
from point $\bf D$ to point $\bf A$. The precession motion continues
on the {\em isochore} where the Hamiltonian becomes constant
(point $\bf A$ to point $\bf B$). In addition, due to dephasing, the amplitude of the precession 
motion is damped. Part of the dephasing is caused by
the energy equilibration with the bath. When pure dephasing is added
the precession motion is damped almost instantly (Cf. fig \ref{fig:traj}).
The motion out of the $\bf \hat B_1$, $\bf \hat B_2$ plane causes
a bending upward in the {\em adiabats} as seen in Fig. \ref{fig:cycle1}.
This bending causes additional work on the {\em adiabats} which is then dissipated
on the {\em isochores}. This causes a reduction in efficiency from 
$\eta_{max}=1-\Omega_a/\Omega_b~=0.5581$ 
which is reached at infinite cycle times to a lower value at maximum power  
$\eta_{Pmax}=0.495$ ($\eta_{carnot}=0.8$).
The cycle of the engine is completely determined by the external control parameters
$\omega_a,\omega_b, T_h,T_c$, and the time allocations: $\tau_h,\tau_c,\tau_{ba},\tau_{ab}$.
Independent of the initial condition, the engine settles to a limit cycle
after a few revolutions with the  preset sequence of {\em isochores} and {\em adiabats}.

The optimization objective is the power of the engine which is the total work
per cycle $\cal W$ divided by the cycle period $\tau$.
Work is obtained only on the {\em adiabats} and is calculated by
integrating the instantaneous power Eq. (\ref{eq:power}) for the {\em adiabat} duration:
${\cal W}_{ba} = \int_0^{\tau_{ba}} {\cal P} dt = \int_0^{\tau_{ba}} \dot \omega
\langle {\bf \hat B_1} \rangle dt$.

The optimization analysis starts by setting the external parameters as the
extreme field values $\omega_b$, $\omega_a$ and the hot and cold bath
temperatures at $T_h$, $T_c$. 
The performance of the engine  therefore will be determined by the time
allocated to the different segments. By setting the total cycle time
$\tau=\tau_{ba}+\tau_c+\tau_{ab}+\tau_h$ the optimization is carried out
by  partitioning the time between the {\it adiabats} and {\it isochores}.
This splits the allocated time between the hot and cold bath {\it isochores}, 
and splits the allocated time between the compression and expansion {\it adiabats}.

In the limiting case of no internal coupling $J=0$, the current model 
is identical to the noted frictionless one \cite{feldmann96,feldmann00}. 
In this frictionless case the optimal
power time allocation on the {\em isochores} becomes: $\Gamma_h \tau_h=\Gamma_c \tau_c$ 
and zero time allocation
on the {\em adiabats}. For $J \ne 0$ the time allocations changes
considerably.
Two limiting cases emerge, the slow limit where most of the cycle
time is allocated to the {\it adiabats}, and the fast or sudden limit
where most of the time is allocated to the {\it isochores}.

Due to the precession motion, the cycle operation is noisy. A small change in 
a parameter can considerably alter the limit cycle and thus 
the performance of the engine. Additional pure dephasing damps the noise 
of the engine (Cf. Fig. \ref{fig:traj}) 
while the overall performance is only slightly altered.

The global power optimum was sought by both a conjugate gradient method and by
a random search scrutinizing local maxima. The optimal power as a function
of the total cycle time  is shown in Fig. \ref{fig:power}
for different $J$ values.
It is clear that the power has a clear maximum with respect to the cycle
period $\tau$. The optimal value decreases and cycle time increases
with increasing $J$.
The maximum power output as a function of $J$ is shown in the insert together
with the analogous friction case.

Despite the large local fluctuations with respect to time allocations
the optimal power performance of the quantum engine shows a remarkable overall similarity
to the performance of an engine subject to phenomenological friction as studied in 
Ref. \cite{feldmann00}. One expects friction to oppose the fast motion on the
{\it adiabats}, therefore the extra power invested 
has to be  independent of the sign of the change in
the control field.  This means that to the lowest order it has to be  proportional
to the square of the time derivative i.e.  ${\cal P}_{fric} = \sigma^2 \dot \omega^2$. 
The accumulated extra work is then dissipated on the cold isochore.  
The engine subject to friction shows performance curves and optimal time allocations
which are very close to the present first principle  quantum model. For this case the origin of
lost power is the inability of the "polarization" vector to follow adiabatically
the instantaneous Hamiltonian. The resulting precession motion on the {\it isochores}
then leads to additional dissipation on the cold {\em isochore}. 
To the lowest order, the additional power should scale as $J^2$ which explains the 
observed linear relation of $J$ with the friction parameter $\sigma$.
\vspace{-.5cm}
\begin{figure}[tb]
\hspace{ 2cm}
\psfig{figure=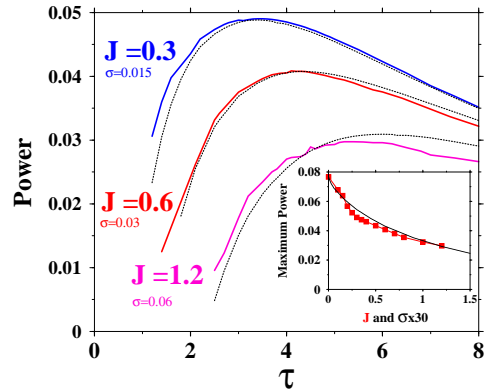,width=0.35\textwidth}
\vspace{-0.2cm}
\caption{Power as a function of cycle time $\tau$ for optimal time allocation
on the four branches for three values of the parameter $J$.
The underlying dotted lines are the optimized power output of an engine
with phenomenological internal friction (in black). A linear relation 
between $J$ and the friction parameter  $\sigma$ fits the data. 
The insert shows the
Maximum power as a function of J (in red) and $\sigma \times 30$.}
\label{fig:power}
\end{figure}
\vspace{-.3cm}
To conclude, we have found that a quantum heat engine with a working fluid
which is not completely controllable by the external field 
shows performance characteristics which can be mapped into
a heat engine subject to  phenomenological friction. 

This research was supported by  the  US Navy under contract 
number N00014-91-J-1498 and the Israel Science Foundation.
The authors wish to thank Jeff Gordon for his continuous help.

\end{document}